\documentclass[ACS,STIX1COL]{WileyNJD-v2}

\articletype{Original article}

\received{27 January 2022}
\revised{2022}
\accepted{2022}

\title{Investigation of the influence of single particle oscillations on the transport properties of dense plasma}

\author[1,2]{S.K. Kodanova}
\author[1,2]{M.K. Issanova*}
\author[2]{G.K. Omiraliyeva}
\author[2]{T.S. Ramazanov}
\authormark{Kodanova S.K. \textsc{et al}}

\address[1]{\orgdiv{Institute of Applied Sciences and IT, 40-48 Shashkin str., 050038 Almaty, Kazakhstan}}
\address[2]{\orgdiv{Institute for Experimental and Theoretical Physics, Al-Farabi Kazakh National University, 71 Al-Farabi ave., 050040 Almaty, Kazakhstan}}

\corres{*M.K. Issanova. \email{issanova@physics.kz }}

\begin{document}
\abstract{Transport properties of the ionic component of dense plasmas  are investigated on the  basis of  effective potentials taking into account the presence of an external alternating electrical (laser) field. The latter generates single particle oscillations of electrons and significantly changes the  screening of the test charge in plasma. Therefore, transport coefficients become non-monotonic function of the oscillation frequency.  The results for the diffusion and viscosity coefficients are presented.
These transport coefficients are computed using the generalized Coulomb logarithm within the Chapman-Enskog model for the fluid description.
It was revealed that the oscillation of the electrons in the alternating external field with the frequency $\omega\gtrsim 2\omega_p$ (with $\omega_p$ being plasma frequency) results in the significant reduction of the  ionic diffusion and viscosity coefficients. In contrast, at $\omega\lesssim \omega_p$, the ionic diffusion and viscosity coefficients are larger compared to the case without the external field impact.
}

\keywords{dense plasma, effective potentials, transport properties, oscillations, plasma frequency}

\jnlcitation{\cname{
\author{Kodanova S.K.}, 
\author{Issanova M.K.}, 
\author{Omiraliyeva G.K.}, 
\author{Ramazanov T.S.}} (\cyear{2022}), 
\ctitle{Influence of the ion core on relaxation processes in dense plasma}, 
\cjournal{Contrib. to Plasma Phys.}, \cvol{2022;00:1--6}.}

\maketitle

\section{Introduction}\label{sec1}

Dense non-ideal plasma physics is relevant for a number of applications such as inertial confinement fusion, laboratory astrophysics, and warm dense matter. 
Therefore, dense non-ideal plasmas have the fundamental importance and relevance for the application in the development of new technology. The latter primarily concerns laser and ion beams induced inertial confinement fusion ~\cite{Hoffmann1, Hoffmann2}. In this regard, the study and quantifying of the transport and dynamic properties of dense plasmas is of high importance  ~\cite{Chen1, Gericke1, Vorberger1}. 

Transport properties of the ionic component of dense plasmas  can be computed directly from molecular dynamics simulation \cite{Hansen}. Other approach is based on the solution of the kinetic equations \cite{Burakovsky1,Kress1}. For that on need to deal with  the collision integral, which contains a logarithmically divergent integral over an impact parameter.
This divergent integral  is commonly replaced by the Coulomb logarithm with the limited  integration range defined by physically motivated arguments, e.g. screening \cite{Ordonez1, Ordonez2}.  Recently, Baalrud and Daligault \cite{Daligault, Baalrud1} have developed the extension of the Coulomb logarithm based approach to the regime of strong coupling with the accuracy close to the molecular dynamics simulations results. 

In previous works, within binary scattering approximation, various transport properties of dense plasmas with weakly degenerate electrons have been investigated using effective (screened) potential approach \cite{Daligault, Baalrud1, Kodanova1, Moldabekov1, Kodanova2, Issanova, MRE18, Ramazanov2021}.  
These investigations were limited to the equilibrium state. However, recently it was shown that the presence of the flux of electrons relative to ions lead to the creation of the wake field, which creates effective ion-ion attraction regions \cite{PhysRevE.91.023102, cpp2015, cpp16}. This phenomena is captured by the use of the dynamically screened dielectric function $\epsilon(k, \Omega=\vec k \cdot \vec v)$ instead of the static version $\epsilon(k, \Omega=0)$, where $\vec v$ is the velocity of the electronic flux relative to an ion. 

Similarly, the presence of the external alternating electrical field (laser field) leads to the break of the ordinary monotonic screening picture of the test charge in plasmas.
This was demonstrated and taken into account for the computation of the electronic transport characteristics by Mulser et al ~\cite{Mulser1, Mulser2}.
Essentially, an external field induces single particle oscillations with the frequency of the eternal field $\omega$. Therefore, screening cannot be described by  the static dielectric function $\epsilon(k, \Omega=0)$, but must be treated using dynamic dielectric function $\epsilon(k, \Omega=\omega)$.
The screened ion potential taking into account the latter fact  was derived by Moldabekov et al \cite{Moldabekov19}.
In this work we extend the investigation of the transport properties of ions in dense plasmas using screened potential by Moldabekov et al \cite{Moldabekov19}. 

To study the transport properties of dense plasma, the potential obtained in Ref. \cite{Moldabekov19} is used in combination with the method for calculating transport properties developed in Refs. \cite{Daligault, Baalrud1}. 
First of all, the screened ion potential is used to compute the  scattering cross sections describing ion-ion binary collisions. Then, the cross sections are used to calculate the generalized Columb logarithm introduced by   Baalrud and Daligault \cite{Daligault, Baalrud1}. This allows to find the ionic diffusion and viscosity coefficients as derived for Chapman-Enskog fluid description. 

We consider two component dense plasma. 
Conventionally,  the plasma state  is characterized by  the coupling parameter $\Gamma=Z^2e^{2}/ak_{B}T_i$ and the density parameter $r_{s}=a_e/a_{B}$, where $Ze$  is the charge of the ion, $k_{B}$ is the Boltzmann constant, $a=(3/4\pi n_i)^{1/3}$ is the average distance between ions (with $n_i$ being the ionic number density), $a_e=(3/4\pi n_e)^{1/3}$ is the mean distance between electrons (with $n_e=n_iZ$),  $a_B$ is the first Bohr radius, and $T_i (T_e)$ is the temperature of ions (electrons).  In this work we consider ionic coupling parameters in the range between $\Gamma=0.1$ and $\Gamma=10$ and density parameter $r_s=1$.



The paper is organized as the following: In Sec.~\ref{sec2} we describe used screened potential; In Sec. ~\ref{sec3} the method used to compute transport properties are presented; In Sec. ~\ref{sec4} we present and discuss the results of the calculations. 

\section{Interaction potential}\label{sec2}

Screening effect due to plasma electrons oscillating in an external electrical field was analyzed in Ref. \cite{Moldabekov19}. 
It was shown how the oscillations of electrons induced by an external field changes the ion-ion interaction in plasmas. When the frequency of the single particle oscillations larger than the electron-plasma frequency $\omega > \omega_{p}$ (where $\omega_{p}=\sqrt{4\pi ne^{2}/m_{e}}$), electrons are not able to efficiently participate in the screening of the test charge. Taking into account this effect the effective potential can be written as \cite{Moldabekov19}  
\begin{equation} \label {potS}
    \Phi(r)= \frac{Z^2e^{2}}{r} \frac{1}{k_{s}^{2}\sqrt{1+\xi^{2}}} \left[\left(A^{2} - \frac{\omega^{2}}{\omega_{p}^{2}}k_{e}^{2}\right)\cos(rA) \right.\\ \left.+  \left(B^{2} + \frac{\omega^{2}}{\omega_{p}^{2}}k_{e}^{2}\right)\exp(-rB)\right], 
\end{equation}
where the following notations are introduced:
\begin{equation} \label {potA}
    A^{2}=\frac{k_{s}^{2}}{2}\left( \sqrt{1+\xi^{2}}-1\right),~~B^{2}=\frac{k_{s}^{2}}{2}\left(\sqrt{1+\xi^{2}}+1\right),~~~\xi=2\frac{\omega}{\omega_{p}} \frac{k_{e} k_{i}}{k_{s}^{2}},
\end{equation}
with the inverse screening length $k_{s}=1/\lambda_{s}$ defined as:  
\begin{equation} \label {kS}
    k_{s}^{2}=k_{e}^{2}\left(1- \frac{\omega^{2}}{\omega_{p}^{2}}\right) + k_{i}^{2} ,
\end{equation}
here $k_{i}^{-1}=\left[(k_{B}T_{i}/4\pi n_{i}e^2 Z^2)+a^2\right]^{1/2}$ \cite{PhysRevE.93.043203} and $k_{e}^2=\frac{1}{2} k_{TF}^2 \theta ^{1/2} I_{-1/2}(\mu/k_BT)$ are the screening length of ions and electrons, where $I_{-1/2}$ denotes the Fermi integral of order $-1/2$ and $\mu$ is the chemical potential of ideal electrons \cite{doi:10.1063/1.4932051, https://doi.org/10.1002/ctpp.202000176}.
Note that  $k_{e}$ interpolates between the Debye and Thomas-Fermi expressions in the limits of classical and strongly degenerate regimes, respectively \cite{doi:10.1063/1.5003910, https://doi.org/10.1002/ctpp.201700113}. $I_{-1/2}(\beta \mu)$ can be computed using a simple interpolation function in terms of the electronic degeneracy parameter $\theta$ as $I_{-1/2}\simeq 4/(4\theta+9\theta^3)^{1/2}$ (see Ref. \cite{PhysRevA.29.1471}), where $\theta=k_BT_e/E_F$ with $E_F$ being the Fermi energy of electrons.
Thus, $k_{s}$ depends on the frequency $\omega$ of the external field and on the plasma parameters.

 If the frequency of the single particle oscillations is less than the electron-plasma frequency $\omega < \omega_{p}$, the equation (\ref{potS})  remains correct if $k_{i}^{2}>k_{e}^{2}\left( 1-\frac{\omega^{2}}{\omega_{p}^{2}}\right)$. Otherwise, if $k_{i}^{2}<k_{e}^{2}\left(1- \frac{\omega^{2}}{\omega_{p}^{2}}\right)$, the potential of the test charge at $\omega > \omega_{p}$ reads \cite{Moldabekov19}:

\begin{equation} \label {pot}
    \Phi(r)= \frac{Z_{i}Z_{j}e^{2}}{r} \frac{1}{k_{s}^{2}\sqrt{1+\xi^{2}}} \left[\left(B^{2} - \frac{\omega^{2}}{\omega_{p}^{2}}k_{e}^{2}\right)\cos(rB)+  \left(A^{2} +\frac{\omega^{2}}{\omega_{p}^{2}}k_{e}^{2} \right)\exp(-rA) \right],   
\end{equation}
where in coefficients $A$ and $B$, the parameter $k_{s}^{2}$ must be replased by $k_{s}^{2}=k_{e}^{2}\left( \frac{\omega^{2}}{\omega_{p}^{2}}-1\right) -k_{i}^{2} $.

The screened potentials (\ref{potS}) and (\ref{pot}) are derived considering weak electron-ion coupling (so that plasma is fully or highly ionized),
 non-relativistic electrons with velocity  $v\ll c$ (where $c$ stands for the light speed), and it is assumed that the electron distribution function is not distorted and there are no instabilities. 
 Moreover, electrons are considered to be thermal meaning that thermalization must be faster than heating.
  This is the case when  $I[{\rm Wcm^{-2}}]\lambda^2[\mu m]<10^{14}$, where  $I$ is the external alternating field intensity, and $\lambda$ is it’s wave length \cite{Atzeni}.
  Additionally, the oscillation of ions in the external field of the laser is neglected since $m_i\gg m_e$.
  


  
  

\section{Transport coefficients}\label{sec3}

In the Chapman-Enskog theory, transport coefficients are defined by the collision integrals \cite{Chapman}. 
The diffusion and viscosity coefﬁcients are determined by collision integrals as the following \cite{Daligault, Baalrud1,  Stanton}:

\begin{equation} \label{Dif1} 
D_{ij} =\frac{3}{16} \frac{k_{B} T}{n\mu_{ij} \Omega _{ij }^{(1,1)} } , ~~~\eta_{ij} =\frac{5}{8} \frac{k_{B} T}{ \Omega _{ij }^{(2,2)} } . 
\end{equation} 
where collision integrals read:

\begin{equation} \label{omega1}
     \Omega_{ij}^{(l,k)} =  \left(\frac{{k_{B} T}}{\pi \mu_{ij}}\right)^{1/2} \int_{0}^{\infty}d\zeta \zeta^{2k+3}e^{-\zeta^2}\sigma_{ij}^{(l)}.
   \end{equation}
with $\mu_{ij}=m_{i}m_{j}/(m_{i}+m_{j})$ being reduced mass and $\zeta^{2} =\mu_{ij}v^{2}/2k_{B} T$.

The $l^{th}$ momentum-transfer cross section is computed using the scattering angle of the binary collision as: 

 \begin{equation} \label {sigma}
     \sigma_{ij}^{(l)} = 2\pi\int_{0}^{\infty}db b  \left[1-\cos^{(l)}\Theta(b,\zeta)\right].
  \end{equation}

The scattering angle is found using \cite{Landau}

\begin{equation} \label {theta}
     \Theta = \pi - 2b \int_{r_{0}}^{\infty}\frac{dr} {r^{2} \sqrt{ 1- \left(\frac{b}{r}\right)^2 - \frac{2}{\mu_{ij}v^{2}} \Phi(r)} }.
\end{equation} 
in which $r_{0}$ is the distance of closest approach, determined from the largest root of the denominator in Eq.(\ref{theta}), $b$ is the impact parameter of the collision, $\Phi(r)$ is the effective interaction potential, which was described in section II. 

Based on Eq. (\ref{omega1}), the Coulomb logarithm associated with the (l,k) collision integral is defined as

\begin{equation} \label {Xi}
 \Xi_{ij}^{(l,k)}= {\frac{1}{2} \int_{0}^{\infty} d \zeta \zeta^{2k+3}e^{-\zeta^2}\sigma_{ij}^{(l)}},
\end{equation}

In the following we present results for the reduced (dimensionless) diffusion and viscosity coefficients:

\begin{equation} \label{Dif2} 
D^{*} = \frac{\sqrt{\pi /3} }{\Gamma ^{5/2} \Xi ^{(1,1)} },~~~\eta^{*} =\frac{5\sqrt{\pi } }{3\sqrt{3} \Gamma ^{{5/ 2}} \Xi ^{(2,2)} }. 
\end{equation} 
where $D^{*}=D/(a^{2} \omega_{p})$ and $\eta^{*}=\eta/(m n \omega_{p} a^{2})$. Note that in Eq. (\ref{Dif2}) we have dropped subscript $ij$ since we consider only one type of ions.  

\section{Results and discussion}\label{sec4}

\subsection{Diffusion coefficients}

\begin{figure}[t]
\centerline{\includegraphics[width=480pt,height=16pc]{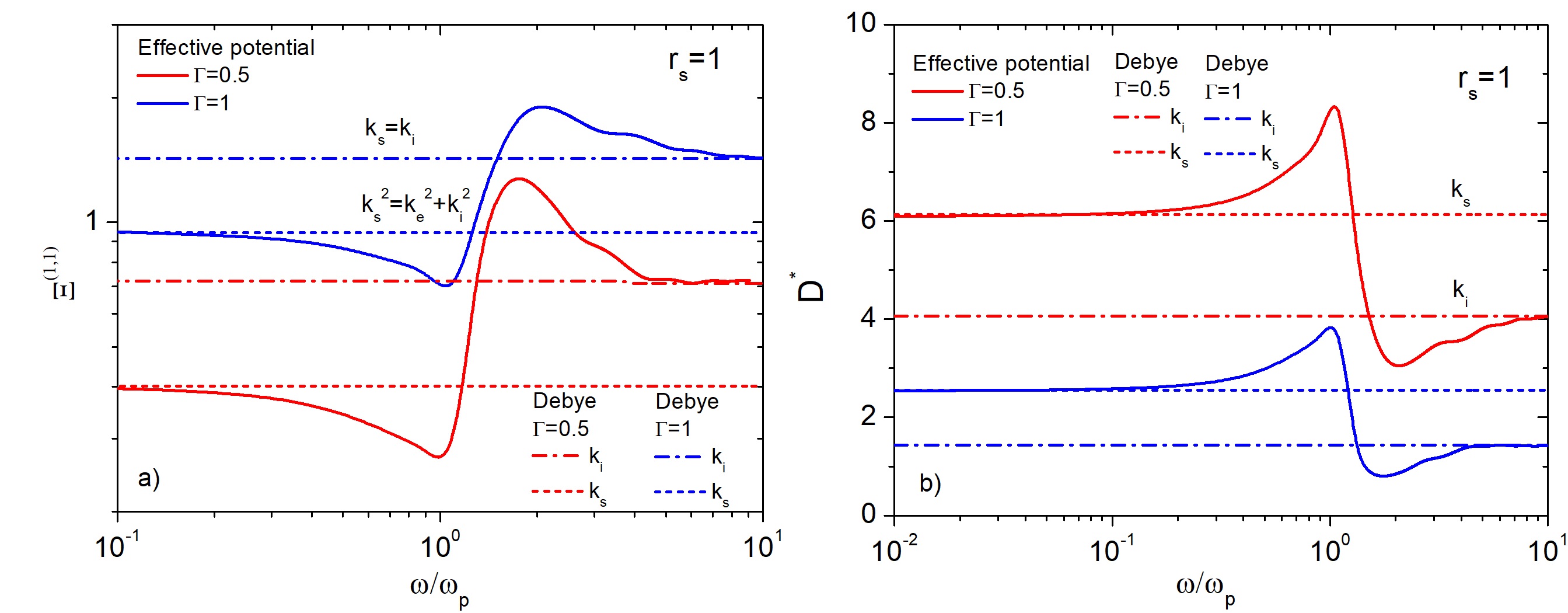}}
\caption{The Coulomb logarithm $\Xi^{(1,1)}$ (a) and diffusion coefficient (b) of ions as the function of $\omega/\omega_{p}$ for different values of $\Gamma$.} 
 \label{fig:1}
\end{figure}

\begin{figure}[t]
\centerline{\includegraphics[width=480pt,height=16pc]{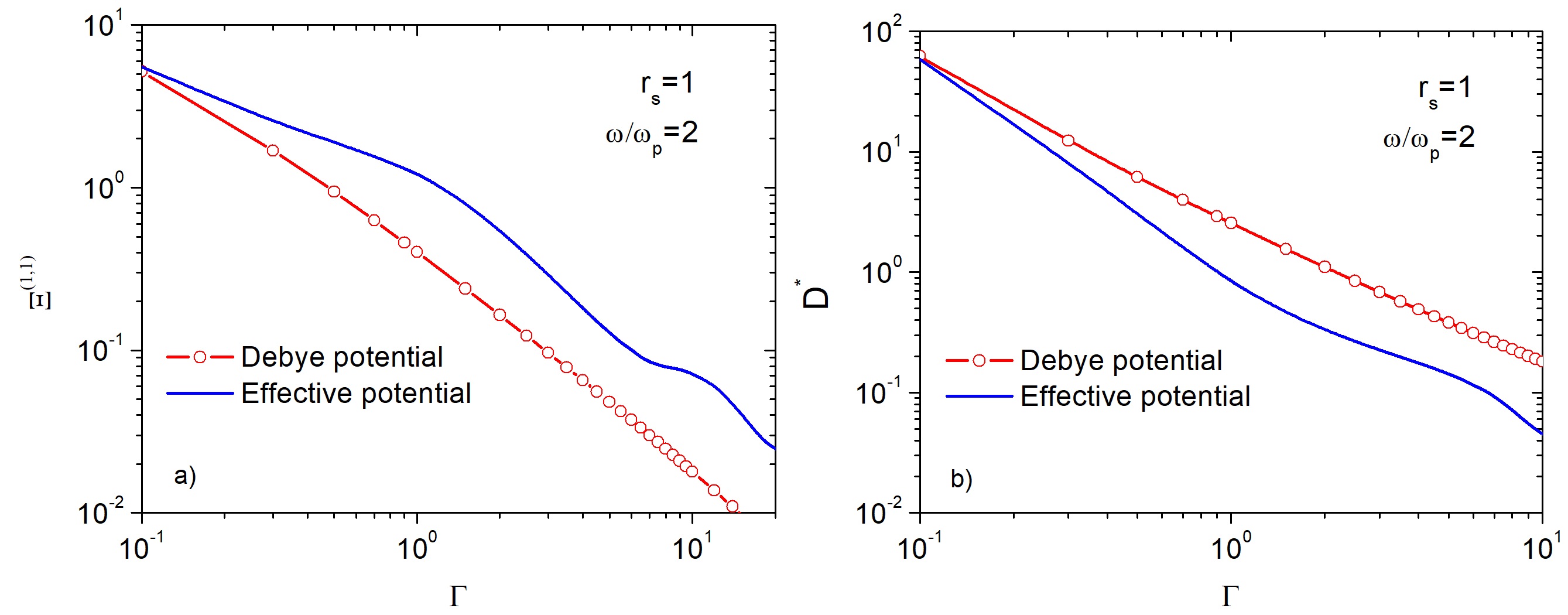}}
\caption{The Coulomb logarithm $\Xi^{(1,1)}$ (a) and diffusion coefficient (b) of ions  as the function of the coupling parameter $\Gamma$ at $\omega/\omega_{p}=2$ and $r_{s}=1$.}
\label{fig:2}
\end{figure}

Fig. \ref{fig:1} a) shows the Coulomb logarithm $\Xi^{(1,1)}$ for the ion-ion interaction as a function of $\omega/\omega_{p}$  in the case $r_{s}=1$ for different values of $\Gamma$.
Red color indicates $\Gamma=0.5$ and blue color corresponds to $\Gamma=1$. 
The solid line corresponds to the effective potential, the dashed and dash-dotted lines correspond to the Debye potential.
In addition to potentials (\ref{potS}) (for $\omega>\omega_p$) and (\ref{pot}) (for $\omega<\omega_p$), we have computed results based on the  Debye potential  with two different 
screening lengths. For the  Debye potential we used screening length $\lambda_{{\rm TCP}}=k_s^{-1}$ of the two component plasmas (TCP) [as defined in Eq. (\ref{kS}) by setting $\omega=0$]  and the screening length $\lambda_{\rm OCP}=k_i^{-1}$ of the one competent plasmas (OCP). In the latter case the screening is due to ions only.
The results for the  Debye potential with $\lambda_{{\rm TCP}}$ is given by dashed lines in Fig. \ref{fig:1}. The results for the  Debye potential with $\lambda_{{\rm OCP}}$ is given by dash-dotted lines in Fig. \ref{fig:1}. By comparing $\lambda_{{\rm TCP}}$ and $\lambda_{{\rm OCP}}$  based data, we see that the Coulomb logarithm is smaller when $\lambda_{{\rm TCP}}$ is used compared to the case when  $\lambda_{{\rm OCP}}$  is used. It is clear that stronger screening in TCP leads to the smaller collision frequency and to the reduction of the Coulomb logarithm compared to OCP model. Now, we look at the results computed using potentials (\ref{potS}) (for $\omega>\omega_p$) and (\ref{pot}) (for $\omega<\omega_p$).
At $\omega\ll \omega_p$, the values of $\Xi^{(1,1)}$ tends to the TCP results computed using the Debye potential. In this case the external field does not disturbe screening cloud around ions and the results are close to the data based on the Debye potentialof the TCP. In contrast, in the limit  $\omega\gg \omega_p$, the values of $\Xi^{(1,1)}$ tends to the results computed using OCP model based on the Debye potential with $\lambda_{{\rm OCP}}$. In this limit, the oscillation of electrons in the field of the external field prevents  electrons from participating in the screening of the ion charge. In between two limits,  $\omega\ll \omega_p$ and $\omega\gg \omega_p$, we observe non-monotonic behavior with 
minimum at $\omega\simeq \omega_p$ and maximum close to $\omega\simeq 2\omega_p$. This is the result of the oscillatory decaying character of effective potentials (\ref{potS})  and (\ref{pot}).

Fig. \ref{fig:1} b) presents the reduced diffusion coefficients as the  function of $\omega/\omega_{p}$  in the case $r_{s}=1$ for different values of $\Gamma$. 
The diffusion coefficients computed using  the Coulomb logarithm $\Xi^{(1,1)}$.  Since $D\sim \frac{1}{\Xi^{(1,1)}}$, the behavior of the diffusion coefficient is explained by the  behavior  the Coulomb logarithm $\Xi^{(1,1)}$. In the limit $\omega\ll \omega_p$, the values of the diffusion coefficient tends from above to the TCP results computed using the Debye potential. 
In the limit  $\omega\gg \omega_p$, the values of the diffusion coefficient tends from bellow to the results computed using the Debye potential with screening length $\lambda_{{\rm OCP}}$.
 The diffusion coefficient has maximum at $\omega\simeq \omega_p$ and minimum close to $\omega\simeq 2\omega_p$. 
The increase in $\Gamma$ from 0.5 to 1.0 results in the decrease in the diffusion coefficient in agreement with earlier studies based on the Debye potential (e.g. see Ref. \cite{doi:10.1063/1.4832016}).
 

In general, an external  field with $\omega<\omega_p$ is not able to penetrate deep into plasma (skin effect).
Therefore, from practical point of view, the case  $\omega>\omega_p$  is more relevant. 
Fig. \ref{fig:2} a) presents the Coulomb logarithm $\Xi^{(1,1)}$ and Fig. \ref{fig:2} b) presents the diffusion coefficient as the function of the coupling parameter $\Gamma$ at $\omega/\omega_{p}=2$ at $r_{s}=1$.
The results are computed using potential (\ref{potS}) and the Debye potential with the screening length  $\lambda_{{\rm TCP}}$.
The solid blue line corresponds to the effective potential (\ref{potS}) and the red circles connected with the solid line correspond to the Debye potential. 
In the considered range of the coupling parameters, the increase in the coupling parameter leads to the decrease of the diffusion coefficient.
From Fig. \ref{fig:2} we see that the oscillation of the electrons induced by the external field can lead to a strong reduction of the diffusion coefficient.
The difference between results computed using potential (\ref{potS}) and the Debye potential increases with the increase in the coupling parameter. At $\Gamma=10$,  potential (\ref{potS}) gives about five times smaller value of the diffusion coefficient compared to the data computed using   the Debye potential.




\subsection{Viscosity coefficients}

\begin{figure}[h]
\centerline{\includegraphics[width=480pt,height=16pc]{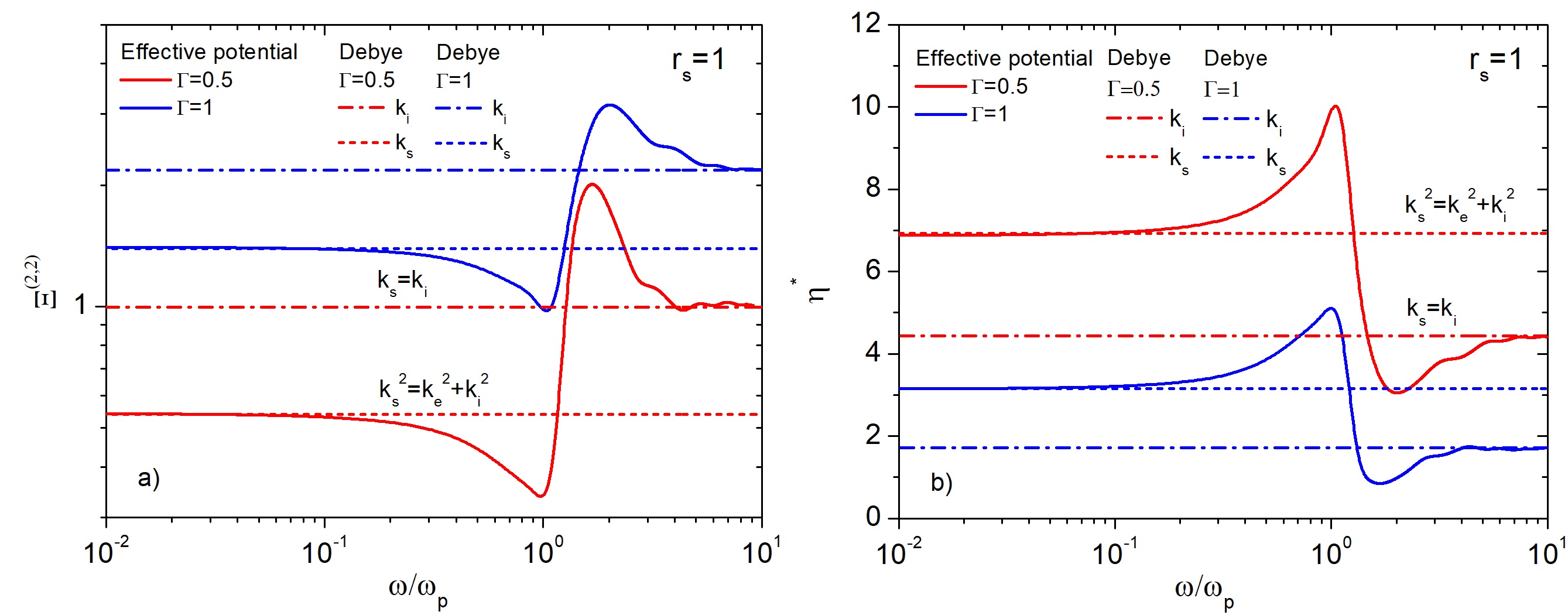}}
\caption{The Coulomb logarithm $\Xi^{(2,2)}$ (a) and viscosity coefficient (b) of ions   as the function of $\omega/\omega_{p}$ for different values of $\Gamma$. }
 \label{fig:3}
\end{figure}

Fig. \ref{fig:3} a) presents the Coulomb logarithm $\Xi^{(2,2)}$ for the ion-ion interaction as the function of $\omega/\omega_{p}$ at $r_{s}=1$ for different values of $\Gamma$. 
Fig. \ref{fig:3} b) shows the viscosity coefficient as the function of $\omega/\omega_{p}$  at $r_{s}=1$ and different values of $\Gamma$.
The solid line corresponds to the effective potential, the dashed and dash-dotted lines correspond to the Debye potential with the screening length $\lambda_{{\rm TCP}}$ and $\lambda_{{\rm OCP}}$, respectively. 

The physics of the behavior of the Coulomb logarithm $\Xi^{(2,2)}$ and the viscosity coefficient with the change in $\omega$ is the same as for the diffusion coefficient.
Both $\Xi^{(2,2)}$ and the viscosity coefficient vary between the Debye potential based TCP and OCP results with the increase of the frequency from $\omega\ll \omega_p$ values to $\omega\gg \omega_p$ regime. A strong non-monotonic behavior takes place between these two limits.  The viscosity coefficient has maximum at $\omega\simeq\omega_p$ and has minimum at  $1.5\omega_p<\omega \lesssim 2\omega_p$.



\begin{figure}[t]
\centerline{\includegraphics[width=480pt,height=16pc]{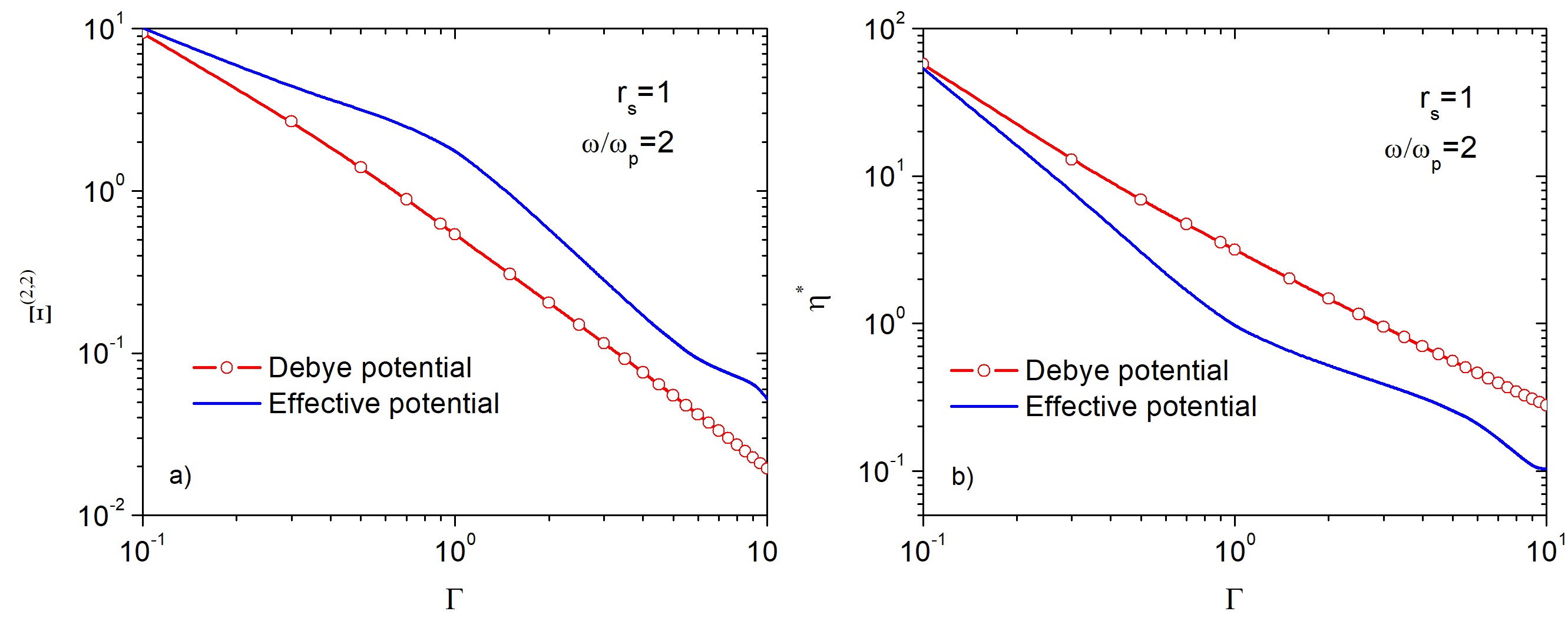}}
\caption{The Coulomb logarithm $\Xi^{(2,2)}$ (a) viscosity coefficient (b) of ions  as the function of the coupling parameter $\Gamma$ at $\omega/\omega_{p}=2$ and $r_{s}=1$.}
\label{fig:4}
\end{figure}

Fig. \ref{fig:4} a) demonstrates the Coulomb logarithm $\Xi^{(2,2)}$ as the  function of the coupling parameter $\Gamma$ at $\omega/\omega_{p}=2$ in the case $r_{s}=1$ for different potentials.  Fig. \ref{fig:4} b) shows the viscosity coefficient as the function of the coupling parameter $\Gamma$ at $\omega/\omega_{p}=2$ and $r_{s}=1$. The solid blue line corresponds to  potential (\ref{potS}) and the red circles connected with the solid line correspond to the Debye potential with screening length $\lambda_{{\rm TCP}}$. 
The increase in $\Gamma$ leads to the decrease (increase) in the viscosity coefficient (the Coulomb logarithm $\Xi^{(2,2)}$).
The difference between the Debye potential based results and the data computed using potential (\ref{potS}) increases with the increase in the coupling parameter.
The results for  the viscosity coefficient  based on potential (\ref{potS}) can be up to 2.5 times smaller  compared to the data calculated using the Debye potential. From figures one can see that at small $\Gamma$ the difference compared to the Debye potential based data is small. At $\Gamma=0.1$ values the influence of the oscillation of electrons in an external field to the diffusion and viscosity coefficients is negligible.

\begin{table}[t]
\caption{The Coulomb logarithm $\Xi^{(1,1)}$ and $\Xi^{(2,2)}$, diffusion and viscosity coefficients of ions  as the function of the coupling parameter $\Gamma$ at $\omega/\omega_{p}=2$ and $r_{s}=1$.}
\label{table01}
\begin{tabular}{ccccccccc}
\hline
    $\Gamma$ & $\Xi^{(1,1)}$ & $D^{*}$ & $\Xi^{(2,2)}$ & $\eta^{*}$ & $\Xi^{(1,1)}_{Debye}$ & $D^{*}_{Debye}$ & $\Xi^{(2,2)}_{Debye}$ & $\eta^{*}_{Debye}$ \\ \hline
    $0.1$ & $5.5329$ & $58.4870$ & $10.0602$ & $53.6110$ & $5.1869$ & $62.3878$ & $21.0243$ & $57.5114$ \\
    $1$ & $1.2124$ & $0.8440$ & $1.7567$ & $0.9708$ & $0.4012$ & $2.5504$ & $0.5394$ & $3.1615$ \\
    $2$ & $0.5457$ & $0.3315$ & $0.5800$ & $0.5197$ & $0.1646$ & $1.0989$ & $0.2045$ & $1.4740$ \\
    $4$ & $0.1819$ & $0.1757$ & $0.1697$ & $0.3139$ & $0.0654$ & $0.4889$ & $0.0760$ & $0.7009$ \\
    $6$ & $0.1013$ & $0.1145$ & $0.0921$ & $0.2098$ & $0.0373$ & $0.3110$ & $0.0419$ & $0.4615$ \\
    $8$ & $0.0790$ & $0.0715$ & $0.0715$ & $0.1317$ & $0.0248$ & $0.2279$ & $0.0272$ & $0.3462$ \\
    $9.5$ & $0.0745$ & $0.0493$ & $0.0598$ & $0.1024$ & $0.0193$ & $0.1901$ & $0.0209$ & $0.2926$ \\
    $12$ & $0.0619$ & $0.0331$ & $0.0324$ & $0.1044$ & $0.0137$ & $0.1493$ & $0.0146$ & $0.2339$ \\
\hline
\end{tabular}
\end{table}



\section{Conclusions}\label{sec5}

We have investigated the effect of the electronic single particle oscillations in the external field on the diffusion and viscosity of the ionic component of dense plasmas.
It is found that when the oscillation frequency is close to  the plasma frequency or smaller than the plasma frequency ($\omega\lesssim \omega_p$), the diffusion and viscosity coefficients are larger than those computed for the equilibrium plasmas. In contrast, the oscillation of the electrons in the alternating external field with the frequency $\omega\gtrsim 2\omega_p$ leads to the significant reduction of the  diffusion and viscosity coefficients  of ions. The results clearly indicate that an external field can significantly modify the diffusion and viscosity characteristics of plasmas. 


\section*{Acknowledgments}
This research was funded by the Science Committee of the Ministry of Education and Science of the Republic of Kazakhstan Grant AP09563273.

\bibliography{wileyNJD-ACS.bib}%

\providecommand{\url}[1]{\texttt{#1}}
\providecommand{\urlprefix}{}
\providecommand{\foreignlanguage}[2]{#2}
\providecommand{\Capitalize}[1]{\uppercase{#1}}
\providecommand{\capitalize}[1]{\expandafter\Capitalize#1}
\providecommand{\bibliographycite}[1]{\cite{#1}}
\providecommand{\bbland}{and}
\providecommand{\bblchap}{chap.}
\providecommand{\bblchapter}{chapter}
\providecommand{\bbletal}{et~al.}
\providecommand{\bbleditors}{editors}
\providecommand{\bbleds}{eds: }
\providecommand{\bbleditor}{editor}
\providecommand{\bbled}{ed.}
\providecommand{\bbledition}{edition}
\providecommand{\bbledn}{ed.}
\providecommand{\bbleidp}{page}
\providecommand{\bbleidpp}{pages}
\providecommand{\bblerratum}{erratum}
\providecommand{\bblin}{in}
\providecommand{\bblmthesis}{Master's thesis}
\providecommand{\bblno}{no.}
\providecommand{\bblnumber}{number}
\providecommand{\bblof}{of}
\providecommand{\bblpage}{page}
\providecommand{\bblpages}{pages}
\providecommand{\bblp}{p}
\providecommand{\bblphdthesis}{Ph.D. thesis}
\providecommand{\bblpp}{pp}
\providecommand{\bbltechrep}{}
\providecommand{\bbltechreport}{Technical Report}
\providecommand{\bblvolume}{volume}
\providecommand{\bblvol}{Vol.}
\providecommand{\bbljan}{January}
\providecommand{\bblfeb}{February}
\providecommand{\bblmar}{March}
\providecommand{\bblapr}{April}
\providecommand{\bblmay}{May}
\providecommand{\bbljun}{June}
\providecommand{\bbljul}{July}
\providecommand{\bblaug}{August}
\providecommand{\bblsep}{September}
\providecommand{\bbloct}{October}
\providecommand{\bblnov}{November}
\providecommand{\bbldec}{December}
\providecommand{\bblfirst}{First}
\providecommand{\bblfirsto}{1st}
\providecommand{\bblsecond}{Second}
\providecommand{\bblsecondo}{2nd}
\providecommand{\bblthird}{Third}
\providecommand{\bblthirdo}{3rd}
\providecommand{\bblfourth}{Fourth}
\providecommand{\bblfourtho}{4th}
\providecommand{\bblfifth}{Fifth}
\providecommand{\bblfiftho}{5th}
\providecommand{\bblst}{st}
\providecommand{\bblnd}{nd}
\providecommand{\bblrd}{rd}
\providecommand{\bblth}{th}
\begin{thebibliography}{1}

\bibitem{Hoffmann1}
D. H. H.~Hoffmann, A.~Blazevic, P.~Ni, {\it {L}aser and {P}article {B}eams} \textbf{2005}, {\it 47}, 23.

\bibitem{Hoffmann2}
D. H. H.~Hoffmann, J.~Jacoby, W.~Laux, {\it {P}hys. {R}ev. {L}ett.} \textbf{1995}, {\it 74}, 1550.

\bibitem{Gericke1}
D. O.~Gericke, M.S.~Murillo, M.~Schlanges, {\it {P}hys. {R}ev. {E} }\textbf{2002}, {\it 65}, 036418.

\bibitem{Chen1}
K.~Chen, S.T.~Sullivan, W.G.~Rellergert, E.R.~Hudson, {\it {O}pt. {E}xpress.} \textbf{2013}, {\it 110}, 173003.

\bibitem{Vorberger1}
J.~Vorberger, D.O.~Gericke, {\it {H}igh {E}nergy {D}ensity {P}hysics} \textbf{2014}, {\it 10}, 1.

\bibitem{Ordonez1}
R. E.~Phillips, C. A.~Ordonez, {\it {AIP} {A}dvances } \textbf{2013}, {\it 3}, 072115.

\bibitem{Ordonez2}
Y.~Chang, C. A.~Ordonez, {\it {P}hys. {R}ev. {E}} \textbf{2000}, {\it 62}, 8564.

\bibitem{Kress1}
J. D.~Kress, J. S.~Cohen, D. A.~Horner, F.~Lambert, L. A.~Collins {\it {P}hys. {R}ev. {E}} \textbf{2010}, {\it 82}, 036404.

\bibitem{Mulser1}
P.~Mulser, M.~Kanapathipillai, {\it {P}hys. {R}ev. {E}} \textbf{2000}, {\it 63}, 016406.

\bibitem{Mulser2}
P.~Mulser, F.~Cornolti, E.~Be´suelle, R.~Schneider, {\it {P}hys. {R}ev. {E}} \textbf{2005}, {\it 71}, 063201.

\bibitem{Kodanova1}
T. S.~Ramazanov, S. K.~Kodanova, Zh. A.~Moldabekov, M. K.~Issanova, {\it {P}hys. {P}lasmas} \textbf{2013}, {\it 20}, 112702.

\bibitem{Moldabekov1}
T. S.~Ramazanov, Zh. A.~Moldabekov, M. T.~Gabdullin, {\it {P}hys. {R}ev. {E}} \textbf{2015}, {\it 92}, 023104.

\bibitem{Ramazanov2021}
T. S.~Ramazanov, S. K.~Kodanova, M. M.~Nurusheva, M. K.~Issanova, {\it {P}hys. {P}lasmas} \textbf{2021}, {\it 28}, 092702.

\bibitem{Kodanova2}
S. K.~Kodanova, T. S.~Ramazanov, M. K.~Issanova, G. N.~Nigmetova, Zh. A.~Moldabekov, {\it {C}ontrib. {P}lasma {P}hys.} \textbf{2015}, {\it 55}, 271.

\bibitem{Issanova}
M. K.~Issanova, S. K.~Kodanova, T. S.~Ramazanov, D. H. H.~Hoffmann, {\it {C}ontrib. {P}lasma {P}hys.} \textbf{2016}, {\it 56}, 425 – 431.

\bibitem{MRE18}
S. K.~Kodanova, M. K.~Issanova, S.M.~Amirov, T. S.~Ramazanov, A.~Tikhonov, Zh. A.~Moldabekov, {\it {M}atter and {R}adiat. at {E}xtremes} \textbf{2018}, {\it 3}, 40-49.

\bibitem{Moldabekov19}
T. S.~Ramazanov, Zh. A.~Moldabekov, M. T.~Gabdullin, {\it {E}ur. {P}hys. {J}. {D}.} \textbf{2018}, {\it 72}, 1-8.

\bibitem{Daligault}
S. D.~Baalrud, J.~Daligault,  {\it {P}hys. {P}lasmas.} \textbf{2014}, {\it 21}, 055707.

\bibitem{Chapman}
S.~Chapman, T.~Cowling,  {\it {C}ambridge {U}niversity} \textbf{1970}.

\bibitem{Baalrud1}
S. D.~Baalrud, J.~Daligault,  {\it {P}hys.{R}ev.{L}ett.} \textbf{2013}, {\it 110}, 235001.

\bibitem{PhysRevE.93.043203}
L.G.~Stanton, M.S.~Murillo, {\it {P}hys.{R}ev.{E}} \textbf{2016}, {\it 93}, 043203.

\bibitem{Stanton}
L.G.~Stanton, M.S.~Murillo, {\it {P}hys.{R}ev.{E}} \textbf{2016}, {\it 93}, 043203.

\bibitem{BaalrudVis1}
J.~Daligault, J.~Rasmussen, S. D.~Baalrud, {\it {P}hys.{R}ev.{E}} \textbf{2014}, {\it 90}, 105.

\bibitem{PhysRevE.91.023102}
Zh.~Moldabekov, P.~Ludwig, M.~Bonitz, T.~Ramazanov, {\it {P}hys.{R}ev.{E}} \textbf{2015}, {\it 91}, 023102.

\bibitem{cpp2015}
Zh.~Moldabekov, P.~Ludwig, J.-P.~Joost, M.~Bonitz, T.~Ramazanov, {\it {C}ontrib. to {P}lasma {P}hys.} \textbf{2015}, {\it 55}, 186-191.

\bibitem{cpp16}
Zh.~Moldabekov, P.~Ludwig, M.~Bonitz, T.~Ramazanov, {\it {C}ontrib. to {P}lasma {P}hys.} \textbf{2016}, {\it 56}, 442-447.

\bibitem{doi:10.1063/1.5003910}
Zh.~Moldabekov, M.~Bonitz, T.~Ramazanov, {\it {P}hys. of {P}lasmas} \textbf{2018}, {\it 25}, 031903.

\bibitem{https://doi.org/10.1002/ctpp.202000176}
Zh.~Moldabekov, T.~Dornheim, T.~Ramazanov, {\it {C}ontrib. to {P}lasma {P}hys.} \textbf{2020}, {\it n/a}, e202000176.

\bibitem{https://doi.org/10.1002/ctpp.201700113}
Zh.~Moldabekov, M.~Bonitz, T.~Ramazanov, {\it {C}ontrib. to {P}lasma {P}hys.} \textbf{2017}, {\it 57}, 499-505.

\bibitem{doi:10.1063/1.4932051}
Zh.~Moldabekov, P.~Ludwig, T.Schoof, M.~Bonitz, T.~Ramazanov, {\it {P}hys. of {P}lasmas} \textbf{2015}, {\it 22}, 102104.

\bibitem{PhysRevB.102.125150}
P.~Hamann, T.~Dornheim, J.~Vorberger, Zh.A.~Moldabekov, M.~Bonitz, {\it {P}hys. {R}ev. {B}} \textbf{2020}, {\it 18}, 125150.

\bibitem{PhysRevA.29.1471}
N.R.~Arista, W.~Brandt, {\it {P}hys. {R}ev. {A}} \textbf{1984}, {\it 29}, 1471-1480.

\bibitem{Atzeni}
S.~Atzeni and J.~Meyer-ter-Vehn, {\it {The Physics of Inertial Fusion}, {O}xford {U}niversityPress, {O}xford} \textbf{2004, 2009}.

\bibitem{Hansen}
J.-P.~Hansen and I.R.~McDonald, {\it {Theory of Simple Liquids}, {A}cademic {P}ress, {O}xford} \textbf{2006}.

\bibitem{doi:10.1063/1.4832016}
K. N.~Dzhumagulova, T.S.~Ramazanov, R.U.~Masheeva, {\it {P}hys. of {P}lasmas} \textbf{2013}, {\it 20}, 113702.

\bibitem{Landau}
L.D.~Landau and E.M.~Lifshitz {\it {Course of theoretical physics}, {B}utterworth-{H}einemann, {O}xford} \textbf{1976}.



\end{thebibliography}

\end{document}